\documentclass[twocolumn,showpacs,preprintnumbers,amsmath,amssymb]{revtex4}

\def\b{\beta}

\def\s{\sigma}

\begin{document}

\title{Universality of REM-like aging in mean field spin glasses}

\author{ G\'erard \surname{Ben Arous}}
\affiliation{Courant Institute, New York University, 251  Mercer St., New
  York, NY 10012-1185}

\author{Anton Bovier}
\affiliation{Weierstrass-Institut f\"ur Angewandte Analysis und Stochastik,
Mohrenstrasse 39, 10117 Berlin, Germany, and\\
Institut f\"ur Mathematik, Technische Universit\"at Berlin, Strasse
des 17.~Juni 136, 10623 Berlin, Germany}

\thanks{Research  supported in part by the DFG in the
  Dutch-German Bilateral Research Group ``Mathematics of random
  spatial models from physics and biology''.}

\author{ Ji\v r\'\i~\v Cern\'y}
\affiliation{ETH Z\"urich, 
  R\"amistrasse 101, 8092 Z\"urich,
Switzerland}

\date{\today}

\begin{abstract}
Aging has become the paradigm to describe dynamical behavior of
glassy systems, and in particular spin glasses. Trap models have been
introduced as simple caricatures of effective dynamics of such
systems. In this Letter we show that in a wide class of mean field
models and on a wide range of time scales, aging occurs precisely as
predicted by the REM-like trap model of Bouchaud and Dean. This is the
first rigorous result about aging in mean field models except for the
REM and the spherical model.
\end{abstract}

\maketitle

A key concept that has emerged over the last years in the study of 
dynamical properties of complex  systems, is that of ``aging''. It is 
applied to systems whose dynamics are dominated by slow transients towards 
equilibrium (see e.g. \cite{BCKM98,FLM01,Biro05,IM05} for  excellent
reviews).
 This phenomena occurs in a huge variety of systems, such as glasses, spin-glasses, 
bio-molecules, polymers, plastics, and has obvious practical implications 
in real-world applications.
 
When discussing aging dynamics, it is all important to specify the precise 
{\it time-scales} considered in relation to the volume. On the one
hand, one may study the  
\textit{non-activated} regime, where the infinite volume limit at 
fixed time $t$ is taken first, and \textit{then} one analyzes the
ensuing dynamics as $t$  
tends to infinity. This \textit{non-activated} regime has been studied
intensively for Langevin dynamics of various soft-spin versions of
mean field spin glasses \cite{CK93,CD95,SCM04,Guio07}. 

For longer time scales, that is times diverging with the volume of the 
system, a full picture is largely missing.  The slow dynamics of complex 
systems in such time scales is often attributed to the presence of 
``thermally activated'' barrier crossings in the configuration space 
\cite{Gold69}. For instance, the standard picture  of the spin glass phase 
typically involves a highly complex landscape of the free energy function 
exhibiting many nested valleys organized according to some hierarchical 
tree structure (see e.g. \cite{BK01,FS02}). To such a picture corresponds 
the heuristic image of a stochastic dynamics that, on time-scales that 
diverge with the size of the system, can be described as performing a 
sequence of ``jumps'' between different valleys at random times those 
rates are governed by the depths of the valleys and the heights of 
connecting passes or saddle points. The extreme situation here
corresponds to considering time scales just before the equilibration
time. While at these scales the relation to the equilibrium Gibbs
distribution is most immediate, in many glassy systems these time
scales appear to be beyond experimental or numerical reach. 

In this letter we show that the mechanism for aging is universal for a
class of Glauber dynamics of $p$-spin Spin Glasses (with $p \ge 3$), out of
equilibrium, in a wide range of time scales. These time scales
are exponentially long but still much shorter than the time scales needed
to reach equilibrium. Thus this mechanism is essentially a transient
one, linked to the exploration of the energy landscape long before the
dynamics can feel the ground state. This mechanism has been first
established for the simple case of the REM, hence our title.  

To capture the  features of activated dynamics, Bouchaud and others 
\cite{Bou92,BD95,MB96,RMB00,BCKM98}  introduced an interesting ansatz, 
that is a mapping of the dynamics onto ``trap models''. These trap models  
are Markov jump processes on a state space that simply enumerates the 
valleys of the free energy landscape. While this picture is intuitively  
appealing, its derivation is based on knowledge obtained in much simpler 
contexts, such as diffusions in finite dimensional potential
landscapes.
Mathematically, trap models are continuous time Markov chains whose
state space is a (infinite or growing with some parameter) graph
(e.g. $\mathbb Z^d$). To the vertices ($=$ traps) of this graph one associates 
independent random variables whose common distribution is assumed to
be heavy tailed, that is their mean is infinite. These variables
represent the \emph{mean waiting times} of the Markov chain in the
corresponding trap. 

Trap models involve the ad-hoc introduction of three major features
that ultimately need justification. This is the independence of the
waiting times associated to the traps, the heavy-tailed nature of
their distribution, and finally the Markov property of the dynamics.

In a series of papers \cite{BC05,BCM06,BC06b,BC07} (see \cite{BC06}
  for a comprehensive review) a systematic 
investigation of a variety of trap models was initiated. In this process,  
it emerged that the "slow-down" of the dynamics appears to be
universal for these trap models (except in the exceptional dimension
1), and,  more precisely, that it has a scaling limit given by  an
$\alpha$-stable 
subordinator. Equivalently it was shown that the classical
Dynkin-Lamperti picture for heavy-tailed renewal processes is
universally satisfied for these trap models (in dimension larger than
1).  

In contrast, very little has been done concerning the derivation of 
trap-model dynamics from stochastic dynamics of even moderately realistic 
spin-glasses, such as the $p$-spin interaction SK models. The only case 
where this has been achieved so far is the simplest of these models, the 
Random Energy Model (REM) of Derrida with a particular form of the 
transition rates. In \cite{BBG02,BBG03,BBG03b} this was achieved by a very 
detailed analysis of the dynamics at time scales just before the 
equilibration time, and at temperatures below the critical one. This 
result relied, in particular, on the detailed understanding of the 
equilibrium distribution of this model. More recently, in \cite{BC06b}, 
the same model was analyzed at much shorter (but still exponentially 
  large) time scales. It emerged that the same aging mechanism is in place 
there and that aging can also occur above the critical temperature.

All these works made crucial use of the independence of energies of
different spin configurations assumed in the definition of the REM. In
the present letter we present the first rigorous aging results in a
model with \emph{correlated energies}, the $p$-spin interaction
Sherrington-Kirkpatrick (SK) model of  spin glasses with $p\geq 3$.  
Quite surprisingly, the results obtained point again to the validity 
of the REM-like trap model as universal aging mechanism. 

\noindent\textit{The $p$-spin SK model.}
We recall that the $p$-spin SK model is defined as follows. A spin 
configuration $\s$ is a vertex of the hypercube $\mathcal S_N\equiv \{-1,1\}^N$. 
The Hamiltonian is given by 
\begin{equation}
  H_N(\s)\equiv 
  -\frac 1{N^{(p-1)/2}}
  \sum_{i_1,\dots,i_p=1}^N
  J_{i_{1}\dots i_p}\s_{i_1}\dots\s_{i_p},
\end{equation}
where $J_{i_1\dots i_p}$ are independent Gaussian random variables
with mean zero and variance one. Alternatively, we can describe the
Hamiltonian as a  centered normal process indexed
by $\mathcal S_N$ with covariance 
\begin{equation}
  \mathbb E[H_N(\sigma )H_N(\tau )]=NR_N(\sigma ,\tau )^p,
\end{equation}
where $R_N(\s,\tau)$ denotes as  usual the 
normalized
overlap, $R_N(\sigma ,\tau )\equiv N^{-1}\sum_{i=1}^N \sigma_i \tau_i$.
We define a random  Gibbs measure on $\mathcal S_N$, 
$\mu_{\b,N}\equiv Z_{\b,N}^{-1}{e^{-\b H_n(\s)}}$.
Note that in the limit $p\uparrow\infty$ one recovers 
the random energy model \cite{GM}, where $H_N(\sigma )$ are 
i.i.d.~Gaussian random variables with variance $N$.

\noindent\textit{Dynamics.}
We consider a continuous time Markov dynamics $\sigma_N (t)$ on 
$\mathcal S_N$ whose transition rates are
\begin{equation}
  \label{e:probas}
  p_{N}(\sigma, \tau )= N^{-1}e^{\beta H_N(\sigma )}
\end{equation}
if $\sigma $ and $\tau $  are related by flipping a single spin, and are
zero otherwise.  It is easy to see that this dynamics is reversible with 
respect to the Gibbs measure $\mu_{\b,N}$. One also sees that it 
represents a nearest-neighbor random walk on the hypercube with traps of 
random depths.

It is thus useful to look at this dynamics as at a time change of a simple 
unbiased discrete-time random walk $Y_N(k)$, $k\in \mathbb N$, on 
$\mathcal S_N$ started \emph{out of equilibrium}\footnote{Note that
  this makes the situation different from the one in
  \cite{Enzo}  where the equilibrated dynamics was
  studied numerically. This makes a comparison of these results with
  ours difficult.}
  at some at some fixed point of $\mathcal S_N$, say at 
$\{1,\dots,1\}$: We define the clock-process by
\begin{equation}
  S_N(k)=
  \sum_{i=0}^{k-1} e_i \exp\big\{-\beta H_N\big(Y_N(i)\big)\big\} ,
\end{equation}
where $(e_i,i\in \mathbb N)$ is a sequence of mean-one i.i.d.~exponential
random variables. Then $\sigma_N(t)$ can be written as
\begin{equation}\label{eq:glauber.1}
  \sigma_N(t)= Y_N(S_N^{-1}(t)).
\end{equation}
$S_N(k)$ is the instant of the $k$-th jump of $\sigma_N(t)$.

\noindent\textit{The REM-like trap model.} 
The idea suggested by the known behavior of the equilibrium 
distribution is that this dynamics, for $\b>\b_c$, will spend long 
periods of 
time in the states $\s^{(1)},\s^{(2)},\dots$ etc. and will move ``quickly''
from one of these configurations to the next. Based on this intuition,
Bouchaud et al. \cite{Bou92,BD95} 
 proposed the ``REM-like'' trap model: Consider a continuous time
Markov process $Z_M$ whose state space is
  the set  $K_M\equiv\{1,\dots,M\}$ of
 $M$ points, representing the $M$ ``deepest'' traps. Each of the 
states is assigned a random 
variable $\varepsilon_k$ (representing minus the energy of the state $k$)
 which is taken to be exponentially
distributed with rate one. 
 If the process is in state $k$, it 
waits an exponentially distributed time with mean proportional 
to $e^{\beta\varepsilon_k}$, and then jumps
with equal probability in one of the other states $k'\in K_M$. 

The  quantity that is used to  characterize the ``aging'' phenomenon is
the  probability $\tilde \Pi_M(t,s)$ that during a time-interval $[t,t+s]$ the 
process does not jump.
 Bouchaud and Dean \cite{BD95} showed that, for $\beta>1$, 
\begin{equation}
  \label{B.8.1}
  \lim_{s,t\uparrow\infty}
  \lim_{M\uparrow\infty}
  \frac {\tilde \Pi_M(t,s)}{H_{1/\beta }(s/t)} =1,
\end{equation}
where the function $H_\alpha $ is defined by 
\begin{equation}
  H_\alpha (w)\equiv \frac {\sin(\pi \alpha )}{\pi} 
\int_{w}^{\infty}   \frac {dx}{(1+x)x^{\alpha}}.
\label{B.53}
\end{equation}

The dynamics of the REM-like trap model can be seen as a time change of a
simple random walk $\tilde Y_M$ on the ``complete graph'' $K_M$ by the clock process,
$\tilde S_M(k)=\sum_{i=0}^{k-1}e_i \exp\{\beta \varepsilon_{Y_M(i)}\}$. As explained
in \cite{BC06b}, the result \eqref{B.8.1} can be deduced from the stronger
claim
\begin{equation}
  \label{e:REMstable}
  \lim_{n\uparrow\infty}
  \lim_{M\uparrow\infty}
  n^{-\beta  } \tilde S_M(n t)= c V_{1/\beta }(t), \qquad t\ge 0,
\end{equation}
where $V_\alpha(t) $ is the  $\alpha $-stable subordinator (increasing
  L\'evy process) with Laplace 
transform 
$\mathbb E[e^{-\lambda V_\alpha (t)}]=\exp(-t\lambda^\alpha )$.

\noindent\textit{The REM.} 
In \cite{BBG02,BBG03,BBG03b} it was confirmed that the 
REM-like picture is correct, at least for the dynamics defined in
\eqref{e:probas}. This result was further extended to shorter time scales
in \cite{BC06b} where the point of view of \eqref{e:REMstable} was put in
the foreground. Namely, it was shown that the clock process converges again
to the stable subordinator: For every $0< \varrho<1$,  
if $\b_\varrho\equiv \b/\sqrt {\varrho} >\b_c\equiv \sqrt{2\ln 2}$, 
$\gamma =\beta \sqrt {2 \varrho \ln 2}$,
\begin{equation}
  \label{e:REMVconv}
  \lim_{N\uparrow \infty}
e^{-\gamma N } N^{\frac{\b_\varrho }{2\beta_c}}
S_N( t  2^{N\varrho})
  =c V_{\beta_c /\beta_\varrho}(t). 
\end{equation}
This implies then a similar aging result as in \eqref{B.8.1}, 
$\Pi (t e^{\gamma N},s e^{\gamma N})\xrightarrow{N\to\infty} 
H_{\beta_c/\beta_\varrho}(s/t)$, 
as in the REM-like trap model for the probability $\Pi_N(t,s)$ that  
$\sigma_N(t)$ does not jump between $t$ and $t+s$.

Note that this result has an interesting interpretation: at a
time-scale 
$e^{\gamma N} N^{-\frac{\b_\varrho }{2\beta_c}} $ 
the process succeeds to make $2^{\varrho N}$
steps, that is it explores a subset of configuration space that
corresponds to a ``little REM'' in volume $n=\varrho N$. At this time 
scale, the process feels an \textit{effective inverse temperature}
$\b_\varrho$.
If the effective temperature is below the critical one for the
standard REM, the system shows aging, otherwise it does not.  It may
seem somewhat counterintuitive that the systems is `effectively
``warming up'' as time goes by. 

Let us discuss the heuristics of this result. When the random walk has
made $2^{\varrho N}$ steps, with $\varrho<1$, it has only explored a
small fraction of the total configuration space. In particular, it has
not had time to find the absolute minima of $H_N$, hence it is still
out of equilibrium. Moreover, the random walk will essentially not
have visited any configuration twice. Therefore, the minimum of $H_N$
along on those configurations that were visited is the minimum of
$2^{\varrho N}$independent Gaussian random variables of mean zero and
variance $N$. It is well know (see e.g. \cite{LLR}) that this is of the
order 
$ N\sqrt {2\varrho\ln 2  }$. Then the mean waiting time in this extreme
trap is of order $\exp(\b N\sqrt{2\varrho\ln 2 })=e^{\gamma N}$, up to a polynomial
correction. Now the condition that $\b_\varrho>\b_c$ implies that this
time is of the same order as the total time the process has
accumulated in all the other sites along its way, and, more precisely,
the process will have spent all but a negligible fraction of its time
in the ``few'' ``deepest trap''. Again standard results of extreme
value theory imply that the precise statistics of the times spent in
the deepest traps are asymptotically governed by a Poisson process, and
that the sum of these random times, after rescaling, converge to a
stable subordinator, as claimed.  

\medskip

\noindent\textit{$p$-spin models} We will now present our main results for the
$p$-spin SK model. The full  
proofs of these results are given  
in \cite{BBC}. First, since the valleys in the free-energy landscape
contains more than one configuration, we should change the two-point
function $\Pi $. We set
\begin{equation}
  \Pi_N^\varepsilon (t,s)
  =\mathbb P
  \{
    R_N\big(
      \sigma_N\big(te^{\gamma N}\big),
      \sigma_N\big((t+s)e^{\gamma N}\big)\big)
    \ge 1-\varepsilon\big\},
\end{equation}
that is the overlap at two far-distant time instants is exceptionally
large.
Then, a similar result as in the REM holds, at least if $p\ge 3$. Namely, 
there is a $p$-dependent constant $\varrho_p$, such that , if  $\varrho$
and $\b$ satisfy the conditions 
\begin{equation}
  \begin{split}
    \label{e:zeta}
    \b_\varrho\equiv \b/\sqrt{\varrho}&> \sqrt {2\ln 2} \equiv \beta_c\\
  \varrho&< \varrho_p, 
  \end{split}
\end{equation}
then, for any $\varepsilon\in(0,1)$, $t>0$, and $s>0$,
\begin{equation}
  \label{e:aging}
  \lim_{N\uparrow \infty }
    \Pi_N^\varepsilon (t,s)=
  H_{\beta_c /\beta_\varrho}(s/t).
\end{equation}
The basis of this result is again the statement analogous to
\eqref{e:REMVconv} that shows that
the properly rescaled clock-process converges to a stable subordinator.

The function $\varrho_p$ used in \eqref{e:zeta} to limit the considered
time scales is increasing and it satisfies 
\begin{equation}
  \label{e:zetabeh}
  \varrho_3\simeq 0.763 
  \qquad \text{and} \qquad
  \lim_{p\uparrow \infty}\varrho_p=1,
\end{equation}
hence in the limit $p\uparrow\infty$ we recover the result for the REM.

Note the r\^ole of the two restrictions on $\b$ and $\varrho$. The
first one is again the statement that the effective temperature at the
time scale considered is below the critical one. The second condition
is related to the correlation of the energies. It implies that the
REM-like behavior holds only up to time scales where the explored
region is still so small that the process does not feel the
correlations; essentially it ensures that the process does not have
enough time to get close enough  to a point it visited before so that
it is able to feel the correlations. 

The heuristic justification of the results in the $p$-spin model is
rather similar to that of the REM. The difference here is that the
energies at the sites that the walk has visited are correlated. Our
assertion is that under the condition $\varrho<\varrho_p$, this has only a
mild effect and does not change the overall picture. The reason for
this relies on the geometric properties of typical trajectories of the
random walk on the hypercube, and on the extreme value properties of
correlated Gaussian processes. First, it has been shown
(see e.g. \cite{Bov06,BGK06,BK08}) that if $p$ is larger than $2$ and if
$K_\varrho$ is a totally random subset of the hypercube $S_N$ of
cardinality $2^{\varrho N}$, with $\varrho$ sufficiently small (depending on
$p$), then the extremal process of $H_N$ restricted to $K_\varrho$ are the
same as if $H_N(\s)$ were independent random variables. 
Note that this is not true in the standard SK model with $p=2$ which
is the reason our results can be expected only for $p\geq 3$.

Now it is
clear that the trajectories of the random walk cannot look exactly
like a totally random set, after all the trajectory is connected,
while in $K_\varrho$ essentially all points are isolated. However, a
detailed analysis of the random walk reveals that its trajectories look
very much like a random set $K_\varrho$ with linear pieces between them
joining the points up in a minimal way. Hence, the correlations have
some impact only very locally in time, implying in particular that deep
traps will not be made of single points  
 but consist of a deep valley
(along the trajectory) that has approximately the same depth and whose
shape and width we can describe quite
precisely. Remarkably, each valley is essentially of a size
independent of $N$ (that is the number of sites contributing
significantly to the residence time in the valley is essentially
finite), and different valleys are statistically independent. 

The fact that traps are finite may appear quite surprising to those 
familiar with the statics of $p$-spin models. From the results there (see 
  \cite{Tal03,Bov06}), one knows that the Gibbs measure concentrates on 
``lumps'' whose radius is of order $N\varepsilon _p$, with 
$\varepsilon_p>0$. The mystery is however solved easily: Around a local minimum
$\sigma_0$ with $H_N(\sigma_0)\sim - \gamma N/\beta $, the process 
$H_N(\s)$ does grows essentially linearly with the distance 
$d(\sigma_0 ,\sigma )$ from the
minimum, 
$\mathbb E[H_N(\sigma )-H_N(\sigma_0)]\sim c(p,\gamma ,\beta )
d(\sigma_0 ,\sigma)$. 
Therefore, the Gibbs mass decreases exponentially with 
$d(\sigma_0 ,\sigma )$. For 
the support of the Gibbs measure, one needs to take into account 
the entropy, that is  that the volumes of the balls of radius $r$ 
increases like $\exp(N(\ln 2-I(1-r/N)))$. For the dynamics, at least at 
our time-scales, this is, however, irrelevant, since the simple random 
walk leaves a local minimum essentially ballistically.

\noindent{\bf Remark: }
We conclude the Letter with a remark on the r\^ole of the 
particular choice of the transition probabilities (\ref{e:probas}) 
depending only on 
starting points. Clearly these favor the proximity to Bouchaud's model. 
For us, on a technical level, the independence of the random walk
trajectory of the random environment defined by the Hamiltonian is
crucial. Even in the case of the REM, we do not know at this point how
to deal with different, and more usual,  
dynamics such as Metropolis or heat bath. 
This problem remains one of the great
challenges in the field.


\begin{thebibliography}{31}
\expandafter\ifx\csname natexlab\endcsname\relax\def\natexlab#1{#1}\fi
\expandafter\ifx\csname bibnamefont\endcsname\relax
  \def\bibnamefont#1{#1}\fi
\expandafter\ifx\csname bibfnamefont\endcsname\relax
  \def\bibfnamefont#1{#1}\fi
\expandafter\ifx\csname citenamefont\endcsname\relax
  \def\citenamefont#1{#1}\fi
\expandafter\ifx\csname url\endcsname\relax
  \def\url#1{\texttt{#1}}\fi
\expandafter\ifx\csname urlprefix\endcsname\relax\def\urlprefix{URL }\fi
\providecommand{\bibinfo}[2]{#2}
\providecommand{\eprint}[2][]{\url{#2}}

\bibitem[{\citenamefont{Bouchaud et~al.}(1998)\citenamefont{Bouchaud,
  Cugliandolo, Kurchan, and M\'ezard}}]{BCKM98}
\bibinfo{author}{\bibfnamefont{J.-P.} \bibnamefont{Bouchaud}},
  \bibinfo{author}{\bibfnamefont{L.}~\bibnamefont{Cugliandolo}},
  \bibinfo{author}{\bibfnamefont{J.}~\bibnamefont{Kurchan}}, \bibnamefont{and}
  \bibinfo{author}{\bibfnamefont{M.}~\bibnamefont{M\'ezard}}, in
  \emph{\bibinfo{booktitle}{Spin glasses and random fields}}, edited by
  \bibinfo{editor}{\bibfnamefont{A.~P.} \bibnamefont{Young}}
  (\bibinfo{publisher}{World Scientific, Singapore}, \bibinfo{year}{1998}).

\bibitem[{\citenamefont{Fisher et~al.}(2001)\citenamefont{Fisher, Le~Doussal,
  and Monthus}}]{FLM01}
\bibinfo{author}{\bibfnamefont{D.~S.} \bibnamefont{Fisher}},
  \bibinfo{author}{\bibfnamefont{P.}~\bibnamefont{Le~Doussal}},
  \bibnamefont{and} \bibinfo{author}{\bibfnamefont{C.}~\bibnamefont{Monthus}},
  \bibinfo{journal}{Phys. Rev. E (3)} \textbf{\bibinfo{volume}{64}},
  \bibinfo{pages}{066107} (\bibinfo{year}{2001}), ISSN
  \bibinfo{issn}{1539-3755}.

\bibitem[{\citenamefont{Biroli}(2005)}]{Biro05}
\bibinfo{author}{\bibfnamefont{G.}~\bibnamefont{Biroli}}, \bibinfo{journal}{J.
  Stat. Mech.} \textbf{\bibinfo{volume}{014}} (\bibinfo{year}{2005}).

\bibitem[{\citenamefont{Igl{\'o}i and Monthus}(2005)}]{IM05}
\bibinfo{author}{\bibfnamefont{F.}~\bibnamefont{Igl{\'o}i}} \bibnamefont{and}
  \bibinfo{author}{\bibfnamefont{C.}~\bibnamefont{Monthus}},
  \bibinfo{journal}{Phys. Rep.} \textbf{\bibinfo{volume}{412}},
  \bibinfo{pages}{227} (\bibinfo{year}{2005}), ISSN \bibinfo{issn}{0370-1573}.

\bibitem[{\citenamefont{Cugliandolo and Kurchan}(1993)}]{CK93}
\bibinfo{author}{\bibfnamefont{L.~F.} \bibnamefont{Cugliandolo}}
  \bibnamefont{and} \bibinfo{author}{\bibfnamefont{J.}~\bibnamefont{Kurchan}},
  \bibinfo{journal}{Phys. Rev. Lett.} \textbf{\bibinfo{volume}{71}},
  \bibinfo{pages}{173} (\bibinfo{year}{1993}).

\bibitem[{\citenamefont{Cugliandolo and Dean}(1995)}]{CD95}
\bibinfo{author}{\bibfnamefont{L.~F.} \bibnamefont{Cugliandolo}}
  \bibnamefont{and} \bibinfo{author}{\bibfnamefont{D.~S.} \bibnamefont{Dean}},
  \bibinfo{journal}{J. Phys. A} \textbf{\bibinfo{volume}{28}},
  \bibinfo{pages}{4213} (\bibinfo{year}{1995}), ISSN \bibinfo{issn}{0305-4470}.

\bibitem[{\citenamefont{Semerjian et~al.}(2004)\citenamefont{Semerjian,
  Cugliandolo, and Montanari}}]{SCM04}
\bibinfo{author}{\bibfnamefont{G.}~\bibnamefont{Semerjian}},
  \bibinfo{author}{\bibfnamefont{L.~F.} \bibnamefont{Cugliandolo}},
  \bibnamefont{and}
  \bibinfo{author}{\bibfnamefont{A.}~\bibnamefont{Montanari}},
  \bibinfo{journal}{J. Statist. Phys.} \textbf{\bibinfo{volume}{115}},
  \bibinfo{pages}{493} (\bibinfo{year}{2004}), ISSN \bibinfo{issn}{0022-4715}.

\bibitem[{\citenamefont{Guionnet}(2007)}]{Guio07}
\bibinfo{author}{\bibfnamefont{A.}~\bibnamefont{Guionnet}}, in
  \emph{\bibinfo{booktitle}{Spin glasses}} (\bibinfo{publisher}{Springer},
  \bibinfo{address}{Berlin}, \bibinfo{year}{2007}), vol. \bibinfo{volume}{1900}
  of \emph{\bibinfo{series}{Lecture Notes in Math.}}, pp.
  \bibinfo{pages}{117--144}.

\bibitem[{\citenamefont{Goldstein}(1969)}]{Gold69}
\bibinfo{author}{\bibfnamefont{M.}~\bibnamefont{Goldstein}},
  \bibinfo{journal}{The Journal of Chemical Physics}
  \textbf{\bibinfo{volume}{51}}, \bibinfo{pages}{3728} (\bibinfo{year}{1969}).

\bibitem[{\citenamefont{Biroli and Kurchan}(2001)}]{BK01}
\bibinfo{author}{\bibfnamefont{G.}~\bibnamefont{Biroli}} \bibnamefont{and}
  \bibinfo{author}{\bibfnamefont{J.}~\bibnamefont{Kurchan}},
  \bibinfo{journal}{Physical Review E} \textbf{\bibinfo{volume}{64}},
  \bibinfo{pages}{016101} (\bibinfo{year}{2001}).

\bibitem[{\citenamefont{Fontanari and Stadler}(2002)}]{FS02}
\bibinfo{author}{\bibfnamefont{J.~F.} \bibnamefont{Fontanari}}
  \bibnamefont{and} \bibinfo{author}{\bibfnamefont{P.~F.}
  \bibnamefont{Stadler}}, \bibinfo{journal}{J. Phys. A.}
  \textbf{\bibinfo{volume}{35}}, \bibinfo{pages}{1509} (\bibinfo{year}{2002}).

\bibitem[{\citenamefont{Bouchaud}(1992)}]{Bou92}
\bibinfo{author}{\bibfnamefont{J.-P.} \bibnamefont{Bouchaud}},
  \bibinfo{journal}{J. Phys. I (France)} \textbf{\bibinfo{volume}{2}},
  \bibinfo{pages}{1705} (\bibinfo{year}{1992}).

\bibitem[{\citenamefont{Bouchaud and Dean}(1995)}]{BD95}
\bibinfo{author}{\bibfnamefont{J.-P.} \bibnamefont{Bouchaud}} \bibnamefont{and}
  \bibinfo{author}{\bibfnamefont{D.~S.} \bibnamefont{Dean}},
  \bibinfo{journal}{J. Phys I(France)} \textbf{\bibinfo{volume}{5}},
  \bibinfo{pages}{265} (\bibinfo{year}{1995}).

\bibitem[{\citenamefont{Monthus and Bouchaud}(1996)}]{MB96}
\bibinfo{author}{\bibfnamefont{C.}~\bibnamefont{Monthus}} \bibnamefont{and}
  \bibinfo{author}{\bibfnamefont{J.-P.} \bibnamefont{Bouchaud}},
  \bibinfo{journal}{J. Phys. A} \textbf{\bibinfo{volume}{29}},
  \bibinfo{pages}{3847} (\bibinfo{year}{1996}).

\bibitem[{\citenamefont{Rinn et~al.}(2000)\citenamefont{Rinn, Maass, and
  Bouchaud}}]{RMB00}
\bibinfo{author}{\bibfnamefont{B.}~\bibnamefont{Rinn}},
  \bibinfo{author}{\bibfnamefont{P.}~\bibnamefont{Maass}}, \bibnamefont{and}
  \bibinfo{author}{\bibfnamefont{J.-P.} \bibnamefont{Bouchaud}},
  \bibinfo{journal}{Phys. Rev. Lett} \textbf{\bibinfo{volume}{84}},
  \bibinfo{pages}{5403} (\bibinfo{year}{2000}).

\bibitem[{\citenamefont{{Be{n A}rous} and {\v C}ern{\'y}}(2005)}]{BC05}
\bibinfo{author}{\bibfnamefont{G.}~\bibnamefont{{Be{n A}rous}}}
  \bibnamefont{and} \bibinfo{author}{\bibfnamefont{J.}~\bibnamefont{{\v
  C}ern{\'y}}}, \bibinfo{journal}{Ann. Appl. Probab.}
  \textbf{\bibinfo{volume}{15}}, \bibinfo{pages}{1161} (\bibinfo{year}{2005}).

\bibitem[{\citenamefont{Be{n A}rous
  et~al.}(2006{\natexlab{a}})\citenamefont{Be{n A}rous, {\v{C}}ern{\'y}, and
  Mountford}}]{BCM06}
\bibinfo{author}{\bibfnamefont{G.}~\bibnamefont{Be{n A}rous}},
  \bibinfo{author}{\bibfnamefont{J.}~\bibnamefont{{\v{C}}ern{\'y}}},
  \bibnamefont{and}
  \bibinfo{author}{\bibfnamefont{T.}~\bibnamefont{Mountford}},
  \bibinfo{journal}{Probab. Theor. Rel. Fields} \textbf{\bibinfo{volume}{134}},
  \bibinfo{pages}{1} (\bibinfo{year}{2006}{\natexlab{a}}).

\bibitem[{\citenamefont{{Be{n A}rous} and {\v C}ern{\'y}}(2008)}]{BC06b}
\bibinfo{author}{\bibfnamefont{G.}~\bibnamefont{{Be{n A}rous}}}
  \bibnamefont{and} \bibinfo{author}{\bibfnamefont{J.}~\bibnamefont{{\v
  C}ern{\'y}}}, \emph{\bibinfo{title}{The arcsine law as a universal aging
  scheme for trap models}}, \bibinfo{howpublished}{to appear in Comm. Pure
  Appl. Math.} (\bibinfo{year}{2008}).

\bibitem[{\citenamefont{{Be{n A}rous} and {\v C}ern{\'y}}(2007)}]{BC07}
\bibinfo{author}{\bibfnamefont{G.}~\bibnamefont{{Be{n A}rous}}}
  \bibnamefont{and} \bibinfo{author}{\bibfnamefont{J.}~\bibnamefont{{\v
  C}ern{\'y}}}, \bibinfo{journal}{Ann. Probab.} \textbf{\bibinfo{volume}{35}},
  \bibinfo{pages}{2356} (\bibinfo{year}{2007}).

\bibitem[{\citenamefont{{Be{n A}rous} and {\v C}ern{\'y}}(2006)}]{BC06}
\bibinfo{author}{\bibfnamefont{G.}~\bibnamefont{{Be{n A}rous}}}
  \bibnamefont{and} \bibinfo{author}{\bibfnamefont{J.}~\bibnamefont{{\v
  C}ern{\'y}}}, in \emph{\bibinfo{booktitle}{{\'E}cole d'{\'Et\'e} de Physique
  des Houches, Session LXXXIII ``Mathematical Statistical Physics''}}
  (\bibinfo{publisher}{Elsevier}, \bibinfo{year}{2006}), pp.
  \bibinfo{pages}{331--394}.

\bibitem[{\citenamefont{Be{n A}rous et~al.}(2002)\citenamefont{Be{n A}rous,
  Bovier, and Gayrard}}]{BBG02}
\bibinfo{author}{\bibfnamefont{G.}~\bibnamefont{Be{n A}rous}},
  \bibinfo{author}{\bibfnamefont{A.}~\bibnamefont{Bovier}}, \bibnamefont{and}
  \bibinfo{author}{\bibfnamefont{V.}~\bibnamefont{Gayrard}},
  \bibinfo{journal}{Phys. Rev. Letts} \textbf{\bibinfo{volume}{88}},
  \bibinfo{pages}{087201} (\bibinfo{year}{2002}).

\bibitem[{\citenamefont{Be{n A}rous
  et~al.}(2003{\natexlab{a}})\citenamefont{Be{n A}rous, Bovier, and
  Gayrard}}]{BBG03}
\bibinfo{author}{\bibfnamefont{G.}~\bibnamefont{Be{n A}rous}},
  \bibinfo{author}{\bibfnamefont{A.}~\bibnamefont{Bovier}}, \bibnamefont{and}
  \bibinfo{author}{\bibfnamefont{V.}~\bibnamefont{Gayrard}},
  \bibinfo{journal}{Comm. Math. Phys.} \textbf{\bibinfo{volume}{235}},
  \bibinfo{pages}{379} (\bibinfo{year}{2003}{\natexlab{a}}).

\bibitem[{\citenamefont{Be{n A}rous
  et~al.}(2003{\natexlab{b}})\citenamefont{Be{n A}rous, Bovier, and
  Gayrard}}]{BBG03b}
\bibinfo{author}{\bibfnamefont{G.}~\bibnamefont{Be{n A}rous}},
  \bibinfo{author}{\bibfnamefont{A.}~\bibnamefont{Bovier}}, \bibnamefont{and}
  \bibinfo{author}{\bibfnamefont{V.}~\bibnamefont{Gayrard}},
  \bibinfo{journal}{Comm. Math. Phys.} \textbf{\bibinfo{volume}{236}},
  \bibinfo{pages}{1} (\bibinfo{year}{2003}{\natexlab{b}}).

\bibitem[{\citenamefont{Gross and Mezard}(1984)}]{GM}
\bibinfo{author}{\bibfnamefont{D.~J.} \bibnamefont{Gross}} \bibnamefont{and}
  \bibinfo{author}{\bibfnamefont{M.}~\bibnamefont{Mezard}},
  \bibinfo{journal}{Nuclear Physics B} \textbf{\bibinfo{volume}{240}},
  \bibinfo{pages}{431} (\bibinfo{year}{1984}).

\bibitem[{\citenamefont{Leadbetter et~al.}(1983)\citenamefont{Leadbetter,
  Lindgren, and Rootz{\'e}n}}]{LLR}
\bibinfo{author}{\bibfnamefont{M.}~\bibnamefont{Leadbetter}},
  \bibinfo{author}{\bibfnamefont{G.}~\bibnamefont{Lindgren}}, \bibnamefont{and}
  \bibinfo{author}{\bibfnamefont{H.}~\bibnamefont{Rootz{\'e}n}},
  \emph{\bibinfo{title}{Extremes and related properties of random sequences and
  processes}}, Springer Series in Statistics
  (\bibinfo{publisher}{Springer-Verlag}, \bibinfo{address}{New York},
  \bibinfo{year}{1983}).

\bibitem[{\citenamefont{Be{n A}rous et~al.}()\citenamefont{Be{n A}rous, Bovier,
  and {\v C}ern\'y}}]{BBC}
\bibinfo{author}{\bibfnamefont{G.}~\bibnamefont{Be{n A}rous}},
  \bibinfo{author}{\bibfnamefont{A.}~\bibnamefont{Bovier}}, \bibnamefont{and}
  \bibinfo{author}{\bibfnamefont{J.}~\bibnamefont{{\v C}ern\'y}},
  \emph{\bibinfo{title}{Universality of the {REM} for dynamics of mean-field
  spin glasses}}, \bibinfo{howpublished}{arXiv:0706.2135}.

\bibitem[{\citenamefont{Bovier}(2006)}]{Bov06}
\bibinfo{author}{\bibfnamefont{A.}~\bibnamefont{Bovier}},
  \emph{\bibinfo{title}{Statistical mechanics of disordered systems}}
  (\bibinfo{publisher}{Cambridge University Press},
  \bibinfo{address}{Cambridge}, \bibinfo{year}{2006}).

\bibitem[{\citenamefont{Be{n A}rous
  et~al.}(2006{\natexlab{b}})\citenamefont{Be{n A}rous, Gayrard, and
  Kuptsov}}]{BGK06}
\bibinfo{author}{\bibfnamefont{G.}~\bibnamefont{Be{n A}rous}},
  \bibinfo{author}{\bibfnamefont{V.}~\bibnamefont{Gayrard}}, \bibnamefont{and}
  \bibinfo{author}{\bibfnamefont{A.}~\bibnamefont{Kuptsov}},
  \emph{\bibinfo{title}{A new {REM} conjecture}},
  \bibinfo{howpublished}{arXiv:math/0612373}
  (\bibinfo{year}{2006}{\natexlab{b}}).

\bibitem[{\citenamefont{Be{n A}rous and Kuptsov}(2008)}]{BK08}
\bibinfo{author}{\bibfnamefont{G.}~\bibnamefont{Be{n A}rous}} \bibnamefont{and}
  \bibinfo{author}{\bibfnamefont{A.}~\bibnamefont{Kuptsov}},
  \emph{\bibinfo{title}{The limits of {REM} universality}},
  \bibinfo{howpublished}{In preparation} (\bibinfo{year}{2008}).

\bibitem[{\citenamefont{Talagrand}(2003)}]{Tal03}
\bibinfo{author}{\bibfnamefont{M.}~\bibnamefont{Talagrand}},
  \emph{\bibinfo{title}{Spin glasses: a challenge for mathematicians}}
  (\bibinfo{publisher}{Springer-Verlag}, \bibinfo{address}{Berlin},
  \bibinfo{year}{2003}).

\bibitem[{\citenamefont{Billoire et~al.}(2005)\citenamefont{Billoire, Giomi,
  and Marinari}}]{Enzo}
\bibinfo{author}{\bibfnamefont{A.}~\bibnamefont{Billoire}},
  \bibinfo{author}{\bibfnamefont{L.}~\bibnamefont{Giomi}}, \bibnamefont{and}
  \bibinfo{author}{\bibfnamefont{E.}~\bibnamefont{Marinari}},
  \bibinfo{journal}{Europhys. Lett.} \textbf{\bibinfo{volume}{71}},
  \bibinfo{pages}{824} (\bibinfo{year}{2005}).

\end{thebibliography}

\def\cprime{$'$}

\end{document}